\documentclass{amsart}

\usepackage{amsmath}
\usepackage{amsthm}
\usepackage{amsfonts}

\def\calP{\mathcal P}
\def\calN{\mathcal N}
\def\R{\mathbb R}
\def\N{\widetilde N}

\def\abs#1{| #1 |}

\let\hat\widehat

\begin{document}

\title{Non-deterministic linear thresholding systems reveal
  their deterministic origins}

\author{Anna Laddach}
\address{Francis Crick Institute, London}

\author{Michael Shapiro}
\address{Francis Crick Institute, London}

\date{22 November 2023}

\maketitle

\section*{Introduction}

Linear thresholding systems have been studied as models of neural
activation \cite{coolen} and more recently as models of cell-intrinsic
gene regulation \cite{ 2017JPhA...50P5601H}.  In these models we have
a set of objects (neurons, genes) which are either on or off (firing
or not, expressed or not).  The model proceeds in discreet time steps.
At each time step, the objects that are on send signals promoting or
repressing other objects.  Each object thresholds the signals it
receives and this determines whether it is on or off at the next time
step.  Thus neurons send excitatory or inhibitory signals to other
neurons; genes promote or inhibit the expression of other genes.
These signals, together with the thresholding operation determine the
next state of the system.  In the non-deterministic case, noise is
added to the signal prior to the thresholding operation.  Here we show
that under mild conditions on the distribution from which random
perturbations are drawn, the underlying deterministic system can be
recovered from the non-deterministic system.

Let us take a moment to formalize this.  If we have $N$ objects, a
{\em state} of the system is a vector of $N$ ones and zeros, i.e., an
element of $x$ of $\{0,1\}^N$ or equivalently, an element of the power
set $\calP = \calP(1,\dots,N)$.  The dynamics of this model involve
the application of a numerical function $f$ to the vector $x$ followed
by thresholding $\theta$ to produce a new vector $y = \theta(f(x)) \in
\{0,1\}^N$.  We will restrict our attention to the case where the
function $f$ is linear, though one can account for interactions
between pairs or triples or larger sets of objects by including
quadratic or higher order terms in $f$.  In addition, we will consider
stochastic versions of these systems by looking at $\theta(f(x) + \N)$
where $\N$ is noise.

The attractors of the deterministic systems are necessarily cycles.
In the cellular metaphor, these cycles are seen as representing stable
cell types.  In the neuronal metaphor, these cycles are seen as
memories \cite{Hopfield}, i.e., a cycle's basin of attraction ``brings
up'' that memory.

\section*{Background and definitions}

Fix $N$.  A {\em state} is a vector $x = (x_1,\dots,x_N) \in
\{0,1\}^N$.  We can think of such a vector as an element of $\calP =
\calP_N$, the power set of $\{1,\dots,N\}$.  State changes according
to a dynamic defined by the pair $(J,\theta)$ where $J$ is an $N\times
N$ matrix and $\theta$ is a thresholding operation.  We take these
thresholds to be positive, so that $0$ is fixed under thresholding.
We abuse notation by using $\theta$ for both the thresholding
function and for the numerical threshhold\footnote{Our notation here
diverges slightly from that of \cite{ 2017JPhA...50P5601H} who use
$\Theta$ as the Heaviside function and include the threshold into its
argument. \cite{ 2017JPhA...50P5601H} also consider separate
thresholds $\theta_i$ for each of the genes before passing to the case
of a single threshold $\theta$.  From a dynamical systems point of
view, this entails no loss of generality as it can be carried out by a
linear change of coordinates for each gene. In particular, one can take
$\theta = \frac{1}{2}$ for every gene.}.  Then $\theta(x_1,\dots,x_N)
= (y_1,\dots,y_N)$ where
$$
y_i = \begin{cases} 
  0 & \text{if  $x_i \le \theta$} \\
  1 & \text{if  $x_i > \theta$} 
\end{cases}
$$
Treating $ x$ as a column vector, we have $f=f_{(J,\theta)}:\calP \to \calP$
by
$$ f_{(J,\theta)}( x) = \theta(J  x)$$

Let $f = f_{(J,\theta)}$ and $f' = f_{(J',\theta')}$.  We will say that $(J,\theta)$
and $(J',\theta')$ are {\em equivalent} if $f:\calP\to\calP$ and
$f':\calP\to\calP$ are the same function and write $(J,\theta) \sim
(J',\theta')$ .

It will be convenient to view the thresholding operation
geometrically.  For $1 \le i \le N$ we take
\begin{align*}
  h_i &= \{(x_1,\dots,x_N) \mid x_i = \theta \} \\
  \hat h &= \cup h_i \\
\end{align*}
The thresholding function is constant on each of the connected
components of $\R^N \setminus \hat h$.  We call these {\em
  compartments} and they biject to $\calP$ under $\theta$.  Thus, for
each $x \in \calP$, we can speak of the compartment $C_x$.

We will say that $(J,\theta)$ is {\em generic} if
$$ J \calP \cap \hat h = \emptyset .$$ We will say that $(J,\theta)$
is {\em robust} if there is an open set $U_J$ around $J$ and an open
set $U_\theta$ around $\theta$ such that for each $J' \in U_J$ and
$\theta' \in U_\theta$, $(J',\theta') \sim (J,\theta)$.  It is not
hard to see that $(J,\theta)$ is robust if and only if it is generic
and that every linear thresholding system is equivalent to a generic
system. 

The function $f$ is a dynamical system on the $2^N$ elements of
$\calP$.  A dynamical system $f$ on a discrete (in this case, finite)
state space can be described as a directed graph.  The vertices $V$ of
this graph are the states.  The edges, $E$ consist of the pairs
$(v,f(v))$ for $v \in V$.  The graph $\Gamma=(V,E)$ necessarily has
the following form: Each connected component of $\Gamma$ consists of a
recurrent cycle of one or more states together with directed trees
where the edges are directed towards the root and the root of each
tree is a vertex of a recurrent cycle.

It's not hard to see that any directed graph $\Gamma$ of the above
form is the graph of a dynamical system $f:V\to V$.  However, not
every such graph is realizable by a function $f= f_{(J,\theta)}$.

We can introduce noise to produce stochastic versions of these
systems.  Suppose we have a measure $\calN$ on $\R^N$.  We will assume
that $\calN$ arises from a density function. Typically, $\calN$ is
taken to be Gaussian or thermal noise.  If $\N$ is a random variable
drawn from this distribution, given $x \in \calP$, we have the random
variable $Y = \theta(Jx + \N)$ which also takes values in $\calP$.  In
this way, for $x,y \in \calP$ we have $p(y \mid x) = p(x \to y) = p(Y
= y)$.  Given $z \in \R^N$ we can define $\calN_z$ to be the probability measure that assigns to any measurable set $U$ the value
$\calN(U - z)$.  With this notation $p(x \to y) = \calN_{Jx}(C_y)$.
The triple $(J,\theta,\calN)$ thus defines a directed labelled graph
$\Gamma = \Gamma_{(J,\theta,\calN)}$ whose edges are pairs $(x,y)$
such that $p(x \to y) > 0$ and each edge is labelled with its
probability.

We now restrict to noise which is symmetric with respect to each of
the coordinates.  We take $\rho_i : \R^N \to \R^N$ to be reflection in
the $i$th coordinate.  That is
 $$ \rho_i(x_1,\dots,x_N) =
(x_1,\dots,x_{i-1},-x_i,x_{i+1},\dots,x_N).$$

We will say that $\calN$ {\em is symmetric} if
\begin{itemize}
\item Every ball around the origin has positive mass, and
\item $\calN$ is symmetric with respect to the $\rho_i$, that is,
  for each $i$ and measurable set $S$, $\calN(S) = \calN(\rho(S))$.  
\end{itemize}

\section*{The non-deterministic system cannot hide its deterministic
  origins}

We are now prepared to state and prove our result.

\medskip \noindent {\bf Theorem}
Suppose $(J,\theta)$ is generic and
that $\calN$ is symmetric.  Suppose $y = f_{(J,\theta)}(x)$ and $y'
\ne y$.  Then for the non-deterministic system $(J,\theta,\calN)$,
$p(y \mid x) > p(y' \mid x)$. In particular,
$\Gamma_{(J,\theta,\calN)}$ determines $\Gamma_{f_{(J,\theta)}}$ as
the subgraph whose edges consist of the maximum probability edge out
of each vertex.
    
\medskip
    
\begin{proof}
 Suppose that $Jx = s = (s_1,\dots,s_N)$

  For $i \in
1,\dots, N$ we take $\delta_i = \abs{\theta -s_i}$.  Since the system is
generic, all of these are positive.  We consider the following
symmetric system of hyperplanes and their complements.  For each $i$,
we set
\begin{align*}
g_i^- &= \{ x \mid x_i = -\delta_i \} \\
g_i^+ &= \{ x \mid x_i = \delta_i \} \\
\hat g &= \cup_i g_i^\pm \\
G_i^{-1}  &= \{ x \mid x_i < -\delta_i \} \\
G_i^{0}  &= \{ x \mid -\delta_i <x_i < \delta_i \} \\
G_i^{1}  &= \{ x \mid \delta_i  < x_i \} \\
\end{align*}

We now consider the transition $x \to y = \theta(s + \N)$.  For each
$i$, thesholding takes place according to one of the hyperplanes $s +
g_i^\pm$. In particular, the value of $y$ depends on which component
of $\R^N \setminus \left(s + \cup g_i^\pm \right)$ the value $s + \N$
lies in, that is, it depends on which component of $\R^N \setminus
\left(\cup g_i^\pm \right)$ $\N$ lies in.  (Since the union of these
hyperplanes has measure 0, we ignore the case where $\N$ lies on one
of them.)  Given a value $y = \theta(s + \N)$, we take the {\em
support} of $y$, $S(y)$, to be the components $C_1,\dots,C_M$ of
$\R^N \setminus \left(\cup g_i^\pm \right)$ such that $y = \theta(s +
\N)$ if and only if $\N \in \cup_{i=1}^M C_i$.  For notational simplicity,
we will assume $y = \theta(s) = 0$ so that in each case the
thresholding hyperplane is $s + g_i^+$.  This does not affect the
argument.

The components of $\R^N \setminus \left(\cup g_i^\pm \right)$ are
indexed by trinary strings of length $N$.  That is to say, each
component has the form
$$ C = \cap_i G_i^{\epsilon_i} $$
where each $\epsilon_i \in \{-1,0,1\}$.  Notice that 
$$ S(0) = \cup_{(\epsilon_1,\dots,\epsilon_N) \in \{-1,0\}^N}
\cap_i G^{\epsilon_i} $$
Observe that for each $i$, the reflection $\rho_i$ fixes the set $\cup
g_i^\pm$, interchanges $G_i^{-1}$ and $G_i^1$ and carries $G_i^0$ to
itself.  By hypothesis, $\rho_i$ preserves $\calN$.  Thus, for each
component $C$ of $\R^N \setminus \left(\cup g_i^\pm \right)$,
$\calN(C) = \calN(\rho_i(C))$.

Suppose, now, that $y' \ne y$.  Let $K$ be the set of indices on which
$y'$ is non-zero.  Let $C_1,\dots,C_M$ be the components that support
$y'$.  These are exactly,
the components whose indices lie in
$$\{(\epsilon_1,\dots,\epsilon_N) \mid
\epsilon_i \in \{-1,0\} \text{ if } i \notin K,
\text{~} \epsilon_i=1 \text{ if } i \in K \}.$$
It now follows that
$$\prod_{k \in K} \rho_k(\cup_{i \in 1,\dots,M} C_i)$$
lies entirely within the support of $y$.  Thus $p(x\to y') \le p(x \to
y)$.  On the other hand, the components of $y'$ do not contain the
component corresponding to $s$, i.e., the component indexed by
$(0,\dots,0)$. By hypothesis, this has positive measure and thus
$p(x\to y') < p(x \to y)$ as required.
\end{proof}

\bibliographystyle{plain}
\bibliography{linearThresholding}

\end{document}